\documentclass[aps,showpacs]{revtex4}
\usepackage{amssymb}
\usepackage{graphicx}

\begin{document}

\title{Scaling theory of magnetism in frustrated Kondo lattices}
\author{V. Yu. Irkhin}
\affiliation{M. N. Mikheev Institute of Metal Physics, 620990 Ekaterinburg, Russia}

\affiliation
{Ural Federal University, 620002 Ekaterinburg, Russia
}

\email{Valentin.Irkhin@imp.uran.ru}


\begin{abstract}
A scaling theory of the Kondo lattices with frustrated exchange interactions is developed,  criterium of antiferromagnetic ordering and quantum-disordered state being investigated. The calculations taking into account magnon and incoherent spin dynamics are performed. Depending on the bare model parameters,  one or two quantum phase transitions into non-magnetic  spin-liquid and Kondo Fermi-liquid ground states can occur with increasing the bare coupling constant. Whereas the renormalization of the magnetic moment in the ordered phase can reach orders of magnitude, spin fluctuation frequency and coupling constant are moderately renormalized in the spin-liquid phase. This justifies application of the scaling approach. Possibility of a non-Fermi-liquid behavior is treated.
\end{abstract}

\maketitle

\section{Introduction}

The Kondo effect (screening of magnetic moments by conduction electrons) is
supposed to be the main issue of anomalous properties observed in a number
of rare earth and actinide systems \cite{Stewart0,Zw,Hewson,IK,Coleman1,Si,Lonzarich,I17}.
Various theoretical methods were developed to treat the competition of the Kondo effect and intersite RKKY coupling, including renormalization group approach \cite{IK} and special mean-field approximations \cite{711,JPCM,Coleman1,Sachdev2,I17}.

On the other hand, magnetic frustrations play a crucial role in strongly correlated systems, in particular in quantum phase transitions \cite{frustr,Vojta}.
There are a number of examples of Kondo systems with such frustrations, including  Pr$_2$Ir$_2$O$_7$ \cite{Nakatsuji}, YbRh$_{2}$Si$_{2}$ \cite{YbRh2Si21}, YbAgGe \cite{Schmiedeshoff}, Yb$_2$Pt$_2$Pb\cite{Aronson}, CePdAl \cite{Zhao1}.

The situation in frustrated correlated metals differs from that in insulating magnets where competition of usual Neel state with valence-bond solid (VBS) phase occurs: instead of VBS, magnetism  competes with the non-magnetic Fermi liquid (FL) state. Here, an intermediate unusual fractionalized Fermi liquid FL$^*$ state with deconfined neutral $S=1/2$ excitations (spinons) can occur which is an analogue of spin liquid \cite{Sachdev1,Sachdev2}.

The Kondo effect reduces the local moment (effective spin) value, thereby
reducing the critical value of frustration necessary for the formation of a
of the quantum disordered state or spin liquid state \cite{711,frustr}.
In turn, the tendency to the  spin-liquid state
(say, Anderson's RVB, the formation of singlets) gives an additional gain for the Kondo state in comparison with magnetically ordered phases. Thus the
magnetic phase diagram becomes complicated, and the problem of quantum phase
transitions occurs: one or two magnetic-nonmagnetic transitions into the
Kondo and/or spin-liquid state take place with increasing the $s-d(f)$ coupling.

As noted in the review \cite{Vojta}, frustrations yield the general tendency
to reduce energy scales, which is manifest in electron or magnon bands of
reduced width and reduced coherence scales, so that essential many-body
renormalizations occur even in the antiferromagnetic (AFM) phase. In the present work, we investigate this problem in the renomalization group approach starting from
weak coupling.

A scaling theory taking into account renormalizations of the effective $%
s-d(f)$ coupling parameter, magnetic moment and spin excitation frequency
was developed in Refs.\cite{IK}. The consideration was performed both in
paramagnetic and magnetically ordered states. However, an important factor
was not taken into account, namely zero-point vibrations which distinguish
ferro- and antiferromagnets and are especially important for frustrated
systems. Moreover, strong singularity of the scaling function in
AFM state resulted in a very narrow critical region where
non-trivial behavior including non-Fermi-liquid (NFL) features occurs.

In the present paper we develop a scaling theory which includes both Kondo
corrections and frustration effects. We well demonstrate that the above
discussed drawbacks are corrected when taking into account zero-point
vibrations. In particular we obtain NFL behavior in magnetic susceptibility
and strong renormalization of the magnetic moment in a rather wide region of
model parameters.

In Sect.2 we discuss the zero-point vibrations in the antiferromagnetic state which give main effect in the Kondo corrections. In Sect.3 we derive the scaling equations in the Kondo model with account of the zero-point vibrations in the equation for sublattice magnetization. The results of numerical solution are presented in Sect.4.
In Sect.5 a non-spin-wave version of spin dynamics is treated.
In Conclusions we discuss possible further developments and experimental implications.

\section{Zero-point vibrations in the Kondo lattice}

We use the Kondo lattice model
\begin{equation}
H=\sum_{\mathbf{k}\sigma }t_{\mathbf{k}}c_{\mathbf{k}\sigma }^{\dagger }c_{%
\mathbf{k}\sigma }-I\sum_{i\sigma \sigma ^{\prime }}\mathbf{S}_{i}%
\mbox {\boldmath
$\sigma $}_{\sigma \sigma ^{\prime }}c_{i\sigma }^{\dagger }c_{i\sigma
^{\prime }}^{{}}+\sum_{ij}J_{ij}\mathbf{S}_{i}\mathbf{S}_{j}  \label{H1}
\end{equation}%
where $t_{\mathbf{k}}$ is the band energy, $\mathbf{S}_{i}$ are spin
operators, $I$ is the $s-d(f)$ exchange parameter, $\sigma $ are the Pauli
matrices.

The intersite exchange interaction $J_{ij}$ includes both a \textquotedblleft
direct\textquotedblright\ exchange (e.g., some kind of superexchange
interaction) and indirect RKKY interaction (to construct perturbation
theory, it is convenient to include the latter in the effective
Hamiltonian). The latter is estimated as $J_{\mathrm{RKKY}}\sim I^{2}\rho $
with $\rho $ being the bare density of states at the Fermi level.



We will treat renormalizations of the staggered magnetic moment $\bar{S}$ and
characteristic spin-wave frequency $\overline{\omega }$ owing to zero-point
vibrations. In frustrated systems they are quite different: $\bar{S}$ is
strongly renormalized, but not $\overline{\omega }$. Moreover, spin wave
description is possible in the disordered phase; the corresponding treatment can be performed within the self-consistent spin wave theory (SSWT) \cite%
{Takahashi,IKK}.


First we present a simple spin-wave consideration.
The sublattice magnetization in the usual spin-wave theory reads
\begin{equation}
\bar{S}=S-\frac{1}{2}\sum_{\mathbf{q}}[S(4J_{\mathbf{Q}}-J_{\mathbf{Q}+%
\mathbf{q}}-J_{\mathbf{Q}-\mathbf{q}}-2J_{\mathbf{q}})/\omega _{\mathbf{q}%
}-1].  \label{eq:6.134}
\end{equation}
with $\mathbf{Q}$ the wavevector of AFM structure.
Then the frustration process may  be treated in terms of softening of the magnon frequency $\omega _{\mathbf{q}}$ (see, e.g., Refs. \cite{IKK,Tamine}).
Following to Ref.\cite{IKK} we can write down at $q\rightarrow 0$ in the two-dimensional (2D) case
\begin{equation}
\omega _{\mathbf{q}}^{2}\simeq S^{2}(J_{\mathbf{Q}}-J_{0})\left[ \alpha
q^{2}+\beta f(\varphi )q^{4}\right] ,  \label{eq:6.135}
\end{equation}%
where  $\beta >0$ and $f(\varphi )$ is a positive polar-angle function of the vector $\bf q$ which is determined by concrete form of the function $J_{\mathbf{q}}$. For $\alpha \rightarrow 0$
we find
\begin{equation}
\bar{S}=S-a\ln {}(\beta /\alpha ),  \label{eq:6.136}
\end{equation}%
\begin{equation}
a=\frac{1}{16\pi ^{2}}(J_{\mathbf{Q}}-J_{0})^{1/2}\int\limits_{0}^{2\pi }%
\frac{1}{[\beta f(\varphi )]^{1/2}}\,d\varphi ,
\end{equation}%
so that $\bar{S}=0$ in some regions of parameters
\begin{equation}
\alpha <\beta \exp {}(-S/a).  \label{eq:6.137}
\end{equation}%
At the same time, zero-point vibrations result in a weak increase of
characteristic spin-wave frequency at not small $q$, for the square lattice $S\rightarrow S+0.079$ in the corresponding prefactor \cite{Takahashi}.

In fact, the condition $\alpha \rightarrow 0$ is not necessary to provide frustration situation, but strong zero-point renormalization of the moment is important for both 2D and 3D cases.
In a more strong treatment, the question about realization of spin liquid in the $J-J^{\prime }$ Heisenberg model for the square and cubic lattices remains open; such a realization is
more probable for triangle-type lattices. However, adding other interactions
can make conditions for this more favorable. In particular, the Kondo
screening reducing effective moment can have such an effect too \cite%
{711}. This influence can be described in terms of decreasing
effective spin \cite{Chakravarty}.
As we shall see below, the softening of the magnon spectrum at small $q$ does not play a crucial role in the Kondo corrections, since these $q$ values do not give dominant contribution (the characteristic quasimomenta are on the Fermi surface).


According to (\ref{eq:6.134}), in the nearest-neighbor approximation for a
Heisenberg antiferromagnet $\bar{S}$ does not depend on the exchange
interaction $J$. On the other hand, such a dependence can be appreciable in a general case,
in particular when approaching the frustration point, e.g., due to
next-nearest-neighbor exchange interaction $J^{\prime }$ being a frustration
parameter. As demonstrate numerical estimations in the self-consistent
spin-wave theory \cite{IKK}, $\partial \bar{S}/\partial |J^{\prime
}/J|=\phi (J^{\prime }/J)$ is about 0.1-1.

We assume that long-range indirect RKKY-interaction $J_{\mathrm{RKKY}}\sim
I^{2}\rho $ results in competition of exchange parameters
and moves the system towards the frustration point, so that $\
|J^{\prime }|\rightarrow \ |J^{\prime }|+tI^{2}\rho $, with $t=\partial
|J^{\prime }|/\partial (I^{2}\rho )>0$ being about unity. Then we have for
the zero-point vibration contribution%
\begin{equation}
(\partial \bar{S}/\partial g)_{zp}=t\phi (J^{\prime
}/J)g/|2J\rho |  \label{zp}
\end{equation}%
with $\ g=-2I\rho $ the dimensionless coupling constant. The factor $%
J\rho $ can be very small (of order of 10$^{-2}$--10$^{-3}$) since the
intersite interaction $J$ is small even with respect to intraatomic
interaction $|I|$.


\section{Scaling equations}



To construct renormalization group equations, we use,
similar to Refs.\cite{IK,I11,731},  the ``poor man
scaling'' approach \cite{And}. In this method one considers
the dependence of effective (renormalized) model parameters on the cutoff
parameter $C\ (-D<C<0)$ which occurs at picking out the Kondo singular
terms. Define the renormalized dimensionless coupling constants%
\begin{equation}
g_{ef}(C)=-2\varrho I_{ef}(C),~g_{ef}(-D)=g
\end{equation}%
The renormalization of $g_{ef}$ in the magnetic state is obtained from renormalization
of the magnetic splitting in electron spectrum.
We find the equation for this by picking up in the sums in the
corresponding self-energies the contribution of intermediate electron states
near the Fermi level with $C<t_{\mathbf{k}}<C+\delta C$:
\begin{equation}
\delta g_{ef}(C)= g_{ef}^{2}\eta (-\overline{\omega }_{ef}/C)\delta C/C
\label{sef}
\end{equation}%
where $\overline{\omega }$ is a characteristic spin-fluctuation energy, $%
\eta (x)$ is the scaling function satisfying the condition $\eta (0)=1$. In
the AFM  phase this reads
\begin{equation}
\eta \left( \overline{\omega }/|C|\right) =\left\langle \left( 1-\omega _{%
\mathbf{k-k}^{\prime }}^{2}/C^{2}\right) ^{-1}\right\rangle
_{t_{k}=t_{k^{\prime }}=E_F}  \label{eta111}
\end{equation}%
where $\left\langle ...\right\rangle $ stands for the average on the Fermi
surface. Unlike Refs. \cite{IK,731}, we use the magnon spectrum of the form
$\omega _{\mathbf{q}}^{2}=c^{2}q^{2}+pq^{4}$. Integration in 2D and 3D cases gives, respectively,

\begin{equation}
\eta ^{AFM}(x,y)=\frac{1}{s_{xy}}\left( \ln \left\vert \frac{%
x^{2}+2y^{2}+s_{xy}}{x^{2}+2y^{2}-s_{xy}}\right\vert -\ln \left\vert \frac{%
x^{2}+s_{xy}}{x^{2}-s_{xy}}\right\vert \right ),  \label{z1}
\end{equation}%

\begin{eqnarray}
\eta ^{AFM}(x,y) &=&\frac{2y^{2}}{s_{xy}}\mathrm{Re}%
([(-x^{2}+s_{xy})(-x^{2}-2y^{2}+s_{xy})]^{-1/2}  \nonumber \\
&&+[(x^{2}+2y^{2}+s_{xy})(x^{2}+s_{xy})]^{-1/2})  \label{z2}
\end{eqnarray}%
with $x=2ck_{F}^{{}}/|C|$, $y=4p^{2}k_{F}^{2}/|C|$, $s_{xy}=\sqrt{%
x^{4}+4y^{2}}$. Putting $\overline{\omega }^{2}=\overline{\omega }%
^{2}(2k_{F})=4c^{2}k_{F}^{2}+16p^{2}k_{F}^{4}\equiv $ $%
4c^{2}k_{F}^{2}(1+r^{2})$ we obtain%
\begin{equation}
\eta ^{AFM}(x)=\eta ^{AFM}\left( \frac{x}{\sqrt{1+r^{2}}},\frac{rx}{\sqrt{%
1+r^{2}}}\right)  \label{z12}.
\end{equation}%
Then the function (\ref{eta111}) has a singularity at $\overline{\omega }=|C|$. Note that at $r=0$ we have ($\theta(x)$ is the step function)
\begin{equation}
\eta ^{AFM}(x)=\left\{
\begin{array}{cc}
{-x^{-2}\ln |1-x^{2}|} & d=3 \\
{(1-x^{2})^{-1/2}\theta (1-x^{2})} & d=2
\end{array}%
\right. \label{intafm}
\end{equation}%
and in the large-$r$ limit we come to the ferromagnetic case (cf. Ref. \cite%
{IK}). To cut the singularities, a small damping with the parameter $\delta $ should be
introduced in the functions (\ref{z1}), (\ref{z2}), see Ref.\cite{IK}.

Besides that, there exists the Kondo contribution which comes from the
incoherent (non-pole) part of the spin spectral density; we will demonstrate
below that it can play an important role.
Then the magnon pole contribution to the spin spectral density is
multiplied by the residue at the pole, $Z=Z(-\overline{\omega }(C)/C)$, and
the incoherent contribution by $1-Z,$
\begin{equation}
\eta (x)=Z(x)\eta ^{AFM}(x)+(1-Z(x))\eta _{incoh}(x) .
\label{incoh}
\end{equation}%
We take for qualitative estimations the scaling function for the paramagnetic case,  $\eta _{incoh}=\eta ^{PM}$,
\begin{equation}
\eta ^{PM}(-\frac{\overline{\omega }}{C})=\int_{-\infty }^{\infty }d\omega
\langle \mathcal{J}_{\mathbf{k-k}^{\prime }}(\omega )(1-\omega
_{{}}^{2}/C^{2})^{-1}\rangle _{t_{k}=t_{k^{\prime }}=E_{F}}  \label{eta11}
\end{equation}%
where $\mathcal{J}_{\mathbf{q}}(\omega )$ is the spectral density of the spin
Green's function, which is normalized to unity,
The integration in the spin-diffusion approximation
\begin{equation}
\mathcal{J}_{\mathbf{q}}(\omega )=\frac{1}{\pi }\frac{\mathcal{D}q^{2}}{%
\omega ^{2}+(\mathcal{D}q^{2})^{2}}  \label{JD}
\end{equation}%
yields \cite{IK}
\begin{equation}
\eta ^{PM}(x)=\left\{
\begin{array}{cc}
\arctan x/x & d=3 \\
\{\frac{1}{2}[1+(1+x^{2})^{1/2}]/(1+x^{2})\}^{1/2} & d=2%
\end{array}%
\right. ,
\end{equation}%
so that at large $x$%
\begin{equation}
\eta ^{PM}(x)=\left\{
\begin{array}{cc}
\pi /2x & d=3 \\
1/\sqrt{2x} & d=2%
\end{array}%
\right.  \label{asymp}
\end{equation}

The Kondo renormalization of spin fluctuation frequency in the paramagnetic phase is determined from the second moment of the spin Green's function
and  in antiferromagnetic phase by calculating the magnon anharmonicity terms \cite{IK,I11,I16}.
Generally speaking, $q^2$ and $q^4$ terms in the magnon spectrum are renormalized in a different way (the situation is similar to the consideration of the anisotropic Kondo model with renormalized gap  \cite{I11}), the corresponding prefactors being determined by some averages over the Fermi surface, which are unknown.
Using the representation $\overline{\omega }_{ef}(C)=\overline{\omega }_{1ef}(C)+\overline{\omega }_{2ef}(C)$ we obtain
\begin{eqnarray}
\delta \overline{\omega }_{1ef}(C)/\overline{\omega}_1 =(a_1/2)g_{ef}^{2}
\eta (-\overline{\omega}_{1ef}/C,-\overline{\omega}_{2ef}/C)\delta C/C  \nonumber \\
\delta \overline{\omega}_{2ef}(C)/\overline{\omega} _2=(a_2/2)g_{ef}^{2}
\eta (-\overline{\omega }_{1ef}/C,-\overline{\omega}_{2ef}/C)\delta C/C.
\label{aa}
\end{eqnarray}
First integral of the system (\ref{sef}), (\ref{aa}) gives
\begin{equation}
\overline{\omega }_{1,2ef}(C)=\overline{\omega}_{1,2}\exp (-(a_{1,2}/2)[g_{ef}(C)-g])
\label{w+g}
\end{equation}
Then we derive
\begin{equation}
\frac{\overline{\omega }_{1ef}(C)}{\overline{\omega}_1 }{\bigg /}
\frac{\overline{\omega }_{2ef}(C)}{\overline{\omega }_2}
=\exp{\bigg (}-\frac{a_1-a_2}{2}[g_{ef}(C)-g] {\bigg )}
\label{w+g12}
\end{equation}
Thus, similar to the renormalization of the gap parameter in the anisotropic model \cite{I11}, we have a  Kondo renormalization of the magnon spectrum. In particular, this can result in the softening of the spectrum at small $q$.
Of course, in a more general treatment, such a softening can play a role in the renormalization process for some parameters,
but this does not play a crucial role since small $q$ do not dominate in the Kondo contributions. Therefore we assume for simplicity that the factors $a_i$ are equal.
For the pure $q^2$-spectrum and  the staggered AFM ordering ($|J|\gg |J^{\prime }|$) we have $a=1-\alpha ^{\prime }$ with%
\begin{equation}
\alpha ^{\prime }\simeq b\frac{J^{\prime }}{J_{{}}}\left\vert \left\langle
\exp (i\mathbf{kR}_{2})\right\rangle _{t_{\mathbf{k}}=E_{F}}\right\vert ^{2}
\end{equation}%
where $b=2$ and $b=4$ for the square and simple cubic lattices, $\mathbf{R}%
_{2}$ runs over the next-nearest neighbors. Thus the difference $1-a$ can be small, so that we put in numerical calculations below $a_{1,2}=a=1$.


%
Then we derive from (\ref{sef}), (\ref{aa})
\begin{eqnarray}
\partial g_{ef}(C)/\partial C &=&-\Lambda  \label{gll} \\
\partial \ln \overline{\omega }_{ef}(C)/\partial C &=&a\Lambda /2,
\label{wll} \\
\partial (1/Z)/\partial C &=&\Lambda /2 , \label{zl}
\end{eqnarray}%
\begin{equation}
\Lambda =[g_{ef}^{2}(C)/C]\eta (-\overline{\omega }_{ef}(C)/C),  \label{l1}
\end{equation}


At numerical calculation it is convenient to use the logarithmic variables
\begin{equation}
\xi =\ln |D/C|,\,\lambda =\ln (D/\overline{%
\omega })\gg 1, \,\Psi (\xi )=\eta (e^{-\xi }).
\end{equation}%
Introducing the function
\begin{equation}
\psi (\xi )=\ln (\overline{\omega }/\overline{\omega }_{ef}(\xi ))
\end{equation}%
which determines renormalization of spin dynamics
we obtain the system of scaling equations%
\begin{equation}
\partial g_{ef}(\xi )/\partial \xi =(2/a)\partial \psi (\xi )/\partial \xi
=g_{ef}^{2}(\xi )\Psi (\lambda +\psi -\xi ).  \label{sc2}
\end{equation}

At the same time, renormalization of $\overline{S}$ is governed by both
Kondo contributions \cite{IK}  and frustrations. Using (\ref{zp}) and (\ref{sc2}) we
obtain
\begin{eqnarray}
\partial \overline{S}(\xi )/\partial \xi  &=&-[\overline{S}+t\phi
(|J^{\prime }(g_{ef}(\xi ))/J|)g_{ef}(\xi )/|2J\rho |]  \nonumber \\
&&\times g_{ef}^{2}(\xi )\Psi (\lambda +\psi -\xi ).  \label{sc221}
\end{eqnarray}%
For simplicity we put in numerical calculations below $\phi =\mathrm{const}$
(linear decrease of $\overline{S}$ with $J'$) so that $t\phi
(|J^{\prime }(g_{ef}(\xi ))/J|)//2|J\rho |\rightarrow A$ and we can estimate
that $A$ can be of order of 10. This large value justifies taking into
account $g^{3}$ terms in the scaling equations.

\section{Results of calculations}


\begin{figure}[h]
  \includegraphics[width=0.45\textwidth]{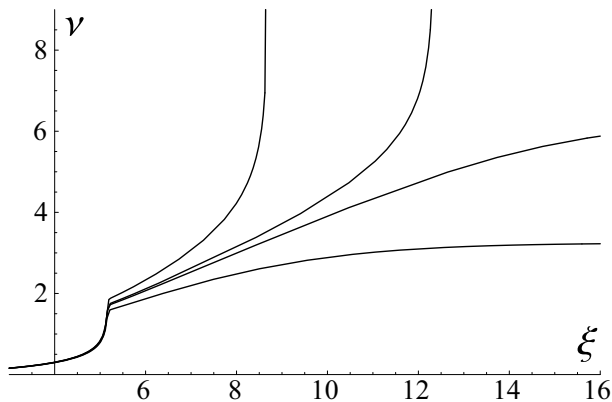}
  \caption{ The behavior of magnetic moment, $\nu (\xi )=\ln (\bar{S}(\xi =0,g)/%
\bar{S}(\xi )),$ for a 2D antiferromagnet in the vicinity of the transition in the spin-liquid state at $g=g_{c1}\simeq  0.1255$. The
parameter values are $\lambda =5$, $\delta =1/100$, $a=1$, $r=0$, $A/\bar{S}%
(\xi =0,g)=8.$ The curves correspond to the coupling constants $g=~$0.125,~0.1255,~0.1256,~0.126 (from below to above).}
\label{1}
\end{figure}

\begin{figure}[h]
  \includegraphics[width=0.45\textwidth]{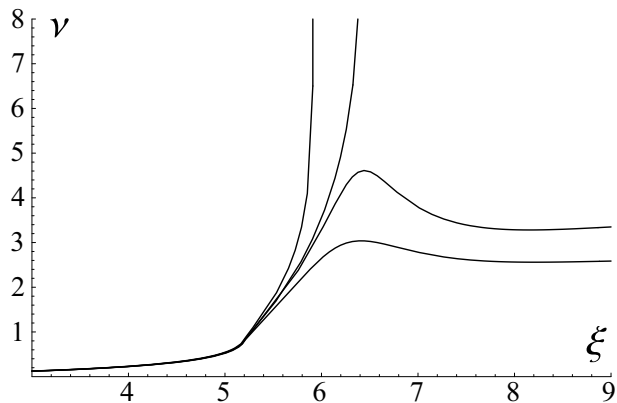}
  \caption{The behavior of magnetic moment
 for a 3D antiferromagnet in the vicinity of $g_{c1}.$ The
parameter values are $\lambda =5$, $\delta =1/100,$ $a=1$, $r=1$, $A/\bar{S}%
(\xi =0,g)=2,$  $g=~$0.133, 0.1334, 0.1335, 0.134 (from below to above).
The maximum in $\nu (\xi )$ is a consequence of the negative minimum in the scaling function.}
\label{2}
\end{figure}

During renormalization process $g_{ef}(\xi )$ increases and $\bar{S}(\xi )$
and $\overline{\omega }_{ef}(\xi )$ decrease. At the first stage $g_{ef}(\xi
)$ follows the one-impurity law, $g_{ef}(\xi )\simeq g/(1-\xi g)$. Then $%
1/g_{ef}(\xi )$ begins to deviate strongly from this behavior, starting from
$\xi \simeq \xi _{1}$ where $\xi _{1}$ is the minimal solution to the
equation
\begin{equation}
\lambda +(a/2)[g_{ef}(\xi )-g]=\xi. \label{soll}
\end{equation}%
If the argument remains negative with further increasing $\xi ,$ $g_{ef}(\xi
)$ will tend to the finite value $g^{\ast },$ the derivative $\partial
(1/g_{ef})/\partial \xi $ being exponentially small, so that scaling almost
stops. For small $g$ we have
\begin{equation}
S^{\ast }=\bar{S}(g(\xi =\infty ))=\bar{S}(g^{\ast }).
\end{equation}%
However, if $g$ is somewhat larger, frustrations start to work at $\xi
\simeq \xi _{1}$, provided that
\begin{equation}
t\phi (|J^{\prime }(g_{ef}(\xi ))/J|)g_{ef}^{3}(\xi _{1})/|2J\rho
|=Ag_{ef}^{3}(\xi _{1})/\bar{S}(\xi _{1})\sim 1.  \label{crit}
\end{equation}%
We rewrite Eq.(\ref{sc221}) as
\begin{equation}
\partial \ln \overline{S}(\xi )/\partial \xi =-[1+Ag_{ef}(\xi )/\overline{S}%
(\xi )]g_{ef}^{2}(\xi )\Psi (\lambda +\psi -\xi ).  \label{new}
\end{equation}%
Then we have the following behavior. Both $\bar{S}(\xi )$ and $\Psi (\lambda
+\psi -\xi )$ become exponentially small at large $\xi $ ($\Psi (-\xi
)\propto \exp (-\xi )$ and $\exp (-\xi /2)$ in the 3D and 2D cases according
to (\ref{asymp})), their ratio being nearly constant, as well as  $%
g_{ef}(\xi )\simeq g^{\ast }.$ Then Eq.(\ref{new}) becomes self-consistent.


Thus we have a linear increase of $\nu (\xi )=\ln (\bar{S}(\xi =0,g)/\bar{S%
}(\xi ))$, $\ \nu (\xi )\simeq \xi -\lambda $ and $\ \nu (\xi )\simeq \xi
/2-\lambda $ respectively, so that%
\begin{equation}
\frac{\bar{S}(C)}{\bar{S}}\propto \left\{
\begin{array}{cc}
|C| & d=3 \\
|C|^{1/2} & d=2%
\end{array}%
\right.
\end{equation}%
The critical value $g_{c1}$ for vanishing of magnetic moment is determined by $%
S^{\ast }(g_{c1})=0$ where

\begin{eqnarray}
\ S^{\ast }(g)=
\bar{S}(J^{\prime }(g^{\ast }(g)))  \nonumber \\
=\bar{S}(g) -\int_{J^{\prime }(g)}^{J^{\prime }(g^{\ast }(g))}d|J^{\prime }(g)/J|\phi
(J^{\prime }/J)  \nonumber \\
=S-\Phi (J^{\prime }(g^{\ast }))+\Phi (J^{\prime }(g))
 \label{gc}
\end{eqnarray}%
with $\Phi (J^{\prime })$ being the contribution of zero-point vibrations
to sublattice magnetization (the Kondo contribution of order of $g^{\ast
2}\rho $ is supposed to be small).

At small $|g-g_{c1}|$, the linear behavior of  $1/ \ln( \bar{S}(\xi ))$ takes place in a
finite, but wide interval of $\xi $. This dependence can be observed as a
NFL behavior in thermodynamic and transport properties ($|C|\rightarrow T$,
cf. Refs. \cite{IK,731,I16}), in particular, in
unusual temperature dependences of magnetic susceptibility $\chi (T) $ in
heavy-fermion systems \cite{Stewart}.

In this regime, we have $S^{\ast }\propto g_{c1}-g, $ unlike small critical
exponents (a situation close to first-order transition) in the
non-frustrated AFM phase \cite{IK}, which are owing to the
singularity of the scaling function.


Thus a two-stage scaling process takes place: first, $g_{ef}(C)$ and $1/\bar{%
S}(\xi )$ increase due to the Kondo effect, and then frustrations lead to
rapid vanishing of $\bar{S}(\xi ).$ Until $g^{\ast }$ is small, the Kondo
term can be neglected. The magnetically disordered state can be treated as a
spin-liquid phase which is characterized by strong spin fluctuations ($\omega ^{\ast }=\overline{\omega }_{ef}(\xi \rightarrow \infty )$ is not small), unlike the Kondo Fermi-liquid (FL) state.

The numerical results on the $\bar{S}(\xi )$ behavior are shown in Figs. 1-3.
The incoherent contribution to the scaling function, which has the asymptotics (\ref{asymp}), is important for the scaling trajectories at large $\xi$.
The account of the $q^{4}$-term ($r\neq 0$) in the magnon spectrum turns out
to be also important for the scaling process in the 3D case where the scaling
function (\ref{z12}), (\ref{z1}) has a negative minimum which dominates over
the incoherent contribution and prevents NFL features for small $r$.
However, the minimum becomes small with increasing $r$, so that we again
obtain appreciable increase of $\ln (1/\bar{S}(\xi ))$, although this
dependence becomes somewhat non-linear. Therefore frustration of the magnon spectrum is important for the NFL behavior.


Most interesting trajectories are obtained at not too large (intermediate) $%
A$ values. The value of $g_{c1}$ weakly depends on $r$, but this dependence is
non-monotonous. Thus the dependence of the scaling trajectories on $r$ turns
out to be non-trivial (Fig. 3).

\begin{figure}[h]
  \includegraphics[width=0.45 \textwidth]{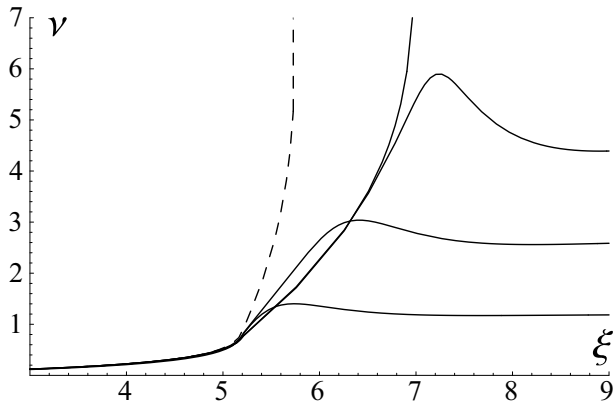}
  \caption{The behavior of magnetic moment, $\nu (\xi )=\ln (\bar{S}(\xi =0,g)/%
\bar{S}(\xi )),$ for a 3D antiferromagnet with $g=0.134\simeq g_{c1}$ and
different $r=$ 0.1, 1, 1.9, 2, 0.3 (from below to above at the right-hand
side of the figure, the line for 0.3 being dashed). Other parameters as in
Fig.2. Note the non-monotonous behavior with $r$: first the maximum in $\nu
(\xi)$ grows with increasing $r,$ so that $\nu (\xi )$ begins to diverge for
$r$ about $0.3$; then the maximum diminishes, but finally it grows again
since $g_{c1}(r)$ falls.}
\label{3}
\end{figure}

For $g>g_{c1},$ two possibilities occur in the scaling behavior with
increasing $\xi $. If $g_{c1}<g<g_{c2},$ the non-magnetic spin liquid
survives as a ground state. With further increasing $g$, starting from $%
g=g_{c2}$, the scaling behavior changes: $(a/2)g_{ef}(\xi )$ increases more
rapidly than $\xi ,$ so that the second solution to (\ref{soll}), $\xi _{2}$,
will occur, and the argument of the function $\Psi $ becomes positive again.
Then $g_{ef}(\xi )$ will diverge at some point $\xi ^{\ast }=\ln
|D/T_{K}^{\ast }|$, $T_{K}^{\ast }$ being the renormalized Kondo
temperature, and $\overline{\omega }_{ef}$ vanishes according to (\ref{w+g}%
). The divergence is described by the law
\begin{equation}
g_{ef}(\xi )\simeq 1/(\xi ^{\ast }-\xi )  \label{div}
\end{equation}%
since $\eta (x\ll 1)=1.$ The behavior (\ref{div}) takes place starting from $%
\xi \simeq \xi _{2}.$ The corresponding numerical results on scaling
trajectories are presented in Figs. 4, 5. A region of critical NFL behavior can occur in $g_{ef}(\xi )$ near $g_{c2},$ but this is very narrow \cite{IK}. Note
that, unlike $\nu (\xi ),$ the maximum in $g_{ef}(\xi )$ is practically not
seen.

\begin{figure}[h]
  \includegraphics[width=0.45 \textwidth]{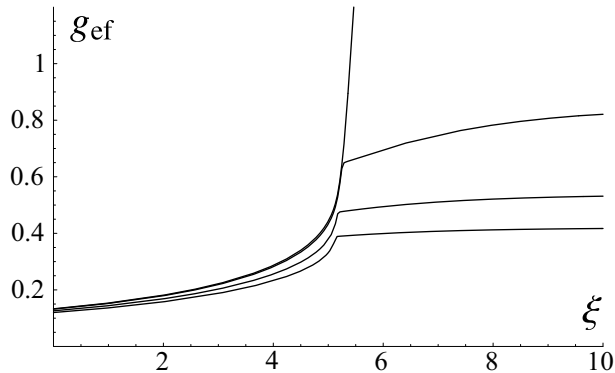}
  \caption{The scaling trajectories $g_{ef}(\xi )$ for a 2D antiferromagnet. The
parameter values are $\lambda =5$, $\delta =1/100,$ $a=1$, $r=0$, $A/\bar{S}%
(\xi =0,g)=8$. The curves  (from below to above) correspond to $g=0.12,$ 0.126 $\simeq g_{c1}$ (the transition point to spin-liquid state), $0.132\simeq g_{c2},$ (the transition point to FL state), 0.133.}
\label{4}
\end{figure}
\begin{figure}[h]
  \includegraphics[width=0.45 \textwidth]{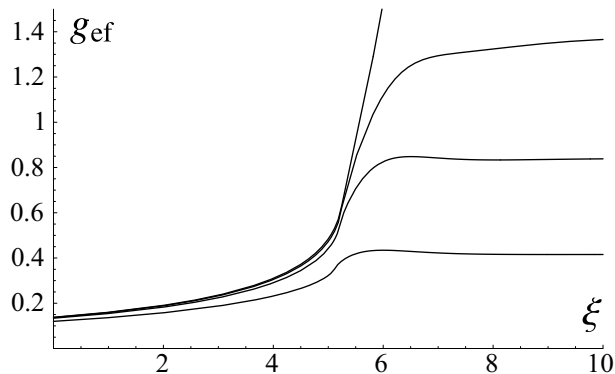}
  \caption{The scaling trajectories $g_{ef}(\xi )$ for a 3D antiferromagnet. The parameter values are $\lambda =5$, $\delta =1/100,$ $a=1$, $r=1$, $A/\bar{S}(\xi =0,g)=2,$ $g=0.12,~0.134\simeq g_{c1},0.137\simeq g_{c2},~0.138$ (from below to above).}
\label{5}
\end{figure}

Thus we should come at $g>g_{c2}$ to a FL ground state with magnetic
fluctuations being suppressed (strictly speaking, this is a crossover into
the strong coupling regime where a unified small energy scale occurs, $%
\omega ^{\ast }\sim T_{K}$ \cite{IK}).






\section{Scaling in the critical regime}

The spin-wave picture in the disordered phase holds in simple theories like SSWT. However, it can be considerably violated in more advanced treatments of quantum phase transitions.
The corresponding excitations acquire incoherent nature and can be described in terms, e.g., spinons -- neutral spin excitations.


At the deconfined quantum critical point or in quantum disordered region the quasiparticle picture is broken and we have the continuum (non-pole) spectral density (see, e.g., Ref. \cite{Senthil3})
\begin{equation}
\mathcal{J}_{\mathbf{q}}(\omega )=-\frac{1}{\pi }{\rm Im}\chi (\mathbf{q}%
,\omega )\sim {\rm Im}\frac{1}{(c^{2}q^{2}-(\omega +i0)^{2})^{1-\eta /2}},
\label{Jeta}
\end{equation}%
where the momentum \textbf{q} is measured from the ($\pi, \pi$) ordering
wavevector of the Neel state, $\eta $ is the anomalous dimension (which is not to be confused with the the scaling function  $\eta (x)$).
Using (\ref{eta11}), (\ref{Jeta}) we obtain for the spin liquid state
\begin{equation}
\eta^{SL}(-\frac{\overline{\omega }}{C})=\left\langle \frac{1}{(1-\omega _{%
\mathbf{k-k}^{\prime }}^{2}/C^{2})^{1-\eta /2}}\right\rangle
_{t_{k}=t_{k^{\prime }}=E_{F}}
\label{int11}
\end{equation}
Large anomalous critical dimension is characteristic for quantum spin liquids
where spin correlations decay algebraically  both at the critical point between the antiferromagnetic and spin-liquid phases and inside the spin-liquid phase, see Ref. \cite{Kaneko} and references therein.
The value of  $\eta=1$ was obtained in Ref. \cite{Senthil3} owing to fractionalized modes within the spinon picture.

In the mean-field approach for $\eta $  for the  Dirac fermions we have the dependence $1/r^{4}$ so that $\eta \simeq 3$.
At the same time, the gauge fluctuations  reduce the mean-field exponent: we have
$1/r^{4-2\alpha }$ with $\alpha \simeq 0.54>1/2$  \cite{Wen,Rantner}.
For the projected d-wave state in the $t-J$ model one obtains
a power-law decay of the equal-time staggered spin-correlation function as
$r^{-\nu }$ with $\nu =1.5$ for the undoped  case and $\nu =2.5$ for 5\% doping  (see review \cite{Wen}); this is considerably slower than the $1/r^{4}$ behavior.


In the 3D case the integration in (\ref{int11})  for the linear spectrum $\omega = cq$ gives
\begin{eqnarray}
\eta^{SL}(x)=\frac{1}{2}\int_{0}^{\pi }\frac{\sin \theta d\theta }{%
[1-x^{2}(1-\cos \theta )]^{1-\eta /2}} \nonumber \\
=\frac{x^{-2}}{\eta /2}[1-(1-x^{2})^{\eta /2}],
\end{eqnarray}%
which agrees with (\ref{intafm}) for $\eta \rightarrow 0$.


In the 2D case the scaling function is expressed in terms of the hypergeometric function $%
_{2}F(a,b,c,z):$%
\begin{eqnarray}
\eta^{SL}(x) &=&\frac{2}{\pi }\int_{0}^{\pi /2}\frac{d\phi }{(1-x^{2}\sin
^{2}\phi )^{1-\eta /2}}  \nonumber \\
&=&\frac{\sin \pi (1-\eta /2)}{\pi ^{3/2}}\left( \frac{1}{\left(
1-x^{2}\right) ^{1/2}}\Gamma (\frac{\eta }{2})\Gamma (\frac{1-\eta }{2}%
)_{2}F_{1}(\frac{1}{2},\frac{\eta }{2},\frac{1+\eta }{2},\frac{1}{1-x^{2}}%
)\right.   \nonumber \\
&&\left. +\frac{1}{\left( 1-x^{2}\right) ^{1-\eta /2}}\Gamma (\frac{\eta -1}{%
2})\Gamma (1-\frac{\eta }{2})_{2}F_{1}(\frac{1}{2},1-\frac{\eta }{2},\frac{%
3-\eta }{2},\frac{1}{1-x^{2}})\right) \theta (1-x^{2})
\label{hyp}
\end{eqnarray}%
with $\Gamma (x)$ the Euler gamma function.
As demonstrate numerical calculations, the expression (\ref{hyp}) is
approximated to high accuracy as
\begin{equation}
\eta^{SL}(x)=\frac{\theta (1-x^{2})}{1-\eta} [ (1-\eta/2)(
1-x^{2})^{\eta/2 -1/2}-\eta/2]
\end{equation}
For $\eta =1$ we have the expression in terms of the elliptic integral
\begin{equation}
\eta^{SL}(x)=\frac{2}{\pi }K(x^{2})\simeq \frac{\theta
(1-x^{2})}{\pi }\ln \frac{16}{1-x^{4}}
\end{equation}%
Thus the scaling function in a 2D spin liquid with $\eta =1$  has the same logarithmic singularity as in the 3D AFM case with standard magnon spectrum.

\begin{figure}[h]
  \includegraphics[width=0.45 \textwidth]{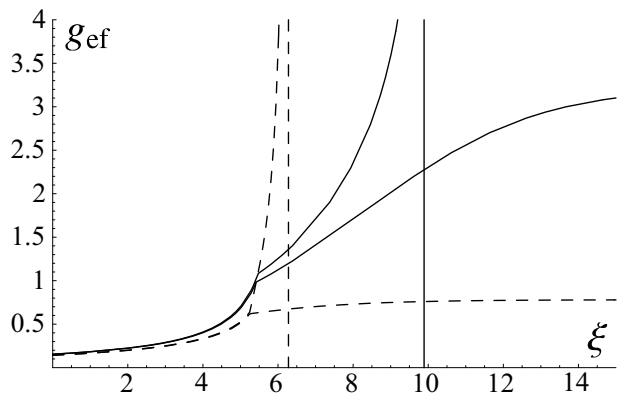}
  \caption{The scaling trajectories $g_{ef}(\xi )$ for a 2D spin-liquid state near $g_c$. The parameter values are $\lambda =5$,  $a=1$. The dashed lines correspond to $\eta=1$,  $g= 0.142, 0.143$, the solid lines to $\eta=1.5$,  $g= 0.154, 0.155$. }
\label{6}
\end{figure}

We may suppose that the scaling function can be again combined from the spinon contribution and electron Kondo-disordered incoherent contribution, so that we can use in numerical calculations the form (\ref{incoh}).
The results are presented in Fig.\ref{6}.
One can see that for small and even moderate $\eta$ values the scaling picture of Fig.\ref{4} is practically not changed.
However, for $\eta \gtrsim  1$ spin-wave-like picture is completely destroyed. Then
the scaling behavior at large $\xi$ is governed  by the paramagnetic contribution, so that we have in a wide region a quasi-NFL behavior which is characterized by increasing of the effective coupling constant and softening of the boson mode \cite{731}.


\section{Conclusions}

In this paper, we have investigated the combined action of magnetic frustrations and Kondo screening in the Kondo lattice model. We have demonstrated that the Kondo effect can lead to considerable enhancing the effect of zero-point vibrations.

The presented scaling approach treats renormalizations of long-range and
short-range magnetic order in a different way. Thereby it enables one to
describe magnetically disordered spin-liquid and Kondo-lattice states in a unified way and discuss possible phase diagrams.

One has to distinguish the quantum transitions and transitions with decreasing $|C|$ (which very roughly correspond to temperature transitions) with a characteristic scale of the renormalized Kondo temperature $T_{K}^{\ast }$ (see discussion in Sect.4).
It should be stressed that the Doniach criterion for magnetic ordering in the Kondo lattices, $g_c\simeq 0.4$ \cite{Don},
which was obtained for a  simplified one-dimensional
model, cannot in fact be quantitatively used for real systems since $g_c$ turn out to be sensitive to parameters of exchange interactions, type of magnetic ordering,
space dimensionality, degeneracy factors  etc. \cite{IK}.
In the present paper, we focus on the strongly frustrated case, where two well-separated quantum transitions occur. The opposite non-frustrated case was investigated in our previous papers \cite{IK,I17} (here, only a narrow NFL region occurs, and for $g>g_c$ a direct transition from the high-temperature paramagnetic phase to the Kondo screened phase takes place).



The first (magnetic) quantum transition with increasing $g$  is determined by vanishing of magnetic ordering due to frustrations and the second transition (strictly speaking, crossover) corresponds to formation of
the non-magnetic Kondo state with screened local moments. It is important
that we do not enter the region of large coupling constant $g_{ef}$ when
treating the first transition. Otherwise, higher order corrections in
scaling equations would make no sense, and we would have only one transition
into the strong coupling region \cite{IK}.

Formally, our scaling gives always two transitions with increasing $g$, since vanishing of $S^{\ast }$ takes place at smaller $g$ than the divergence of $g^{\ast }.$
However, the difference becomes practically non-observable for small
frustration. The boundary of these two regimes is determined by the
criterium (\ref{crit}). In a sense, the situations with two phase
transitions is similar to that in itinerant magnets where local moment
formation and magnetic ordering can be separated.

As for  transitions with decreasing  $|C|$ or temperature, at $g_{c1}<g<g_{c2}$ they go into a spin-liquid-type state and at $g>g_{c2}$ directly to FL state (in the latter case, the incoherent scaling function should be used \cite{IK}). This picture is similar to that of Refs.\cite{Vojta,Sachdev1,Sachdev2} treating the critical non-Fermi-liquid FL$^*$ phase which is an analogue of  spin-liquid state. However, occurrence of a (secondary) itinerant magnetic ordering in the low-temperature (strong coupling) regime is not excluded, as discussed in \cite{Sachdev1,Sachdev2}.
This is in agreement with experimental low-temperature magnetic phase diagrams of the frustrated heavy-fermion and Kondo lattice systems, which are rather complicated.
Thus the problem of stability of the metallic spin-liquid in the ground state (competing with magnetically ordered and VBS states) requires further investigations.

There are a number of Kondo systems demonstrating frustrated magnetism \cite{Nakatsuji,Schmiedeshoff,Aronson}.
In particular, one should  mention recent results on the distorted Kagome intermetallic system CePdAl
where magnetic frustration of 4f-moments gives rise to  a paramagnetic quantum-critical phase \cite{Zhao1,Zhao}.  The picture proposed in \cite{Zhao} includes also f-electron delocalization (transition from small to large Fermi surface), considered in Ref.\cite{Sachdev2}.

To treat concrete experimental situations, calculations with realistic
magnon spectrum would be instructive. It should be noted that 2D-like spin
fluctuations are typical not only for cuprates: two-dimensional fluctuations
at the quantum-critical point are observed in CePdAl  \cite{Zhao1},
CeCu$_{6-x}$Au$_{x}$ \cite{2D}
and YbRh$_{2}$Si$_{2}$ \cite{YbRh2Si2}. The picture presented in the present work can be used to obtain information about frustration degree and magnon spectrum in real
anomalous $f$-systems. The softening of the spectrum owing to the Kondo renormalization given by (\ref{w+g12}) should be also mentioned.


I am grateful to Yu.~N.~Skryabin and M.I. Katsnelson for helpful
discussions.
The research was carried out within the Ministry State Assignment  of Russia (theme ``Quantum'' No. AAAA-A18-118020190095-4) and partially supported by Program 211 of the Government of the Russian Federation (Agreement 02.A03.21.0006).

\end{document}